\documentclass[twocolumn]{aastex62}

\usepackage{subfigure}
\usepackage{amsmath}
\usepackage{multirow}
\usepackage{natbib}

\shorttitle{SED modelling of Orphan $\gamma$-ray flares} 
\shortauthors{Patel et al.}

\begin{document}


\title{Broadband Modelling of Orphan Gamma Ray Flares}

\correspondingauthor{S. R. Patel}
\email{sonalpatel.982@gmail.com}

\author{S. R. Patel}
\affiliation{Department of Physics, University of Mumbai, Santacruz (E), Mumbai-400098, India}

\author{D. Bose}
\affiliation{Department of Physics, Indian Institute of Technology Kharagpur, Kharagpur-721302, India}

\author{N. Gupta}
\affiliation{Raman Research Institute, C. V. Raman Avenue, Sadashivanagar, Bangalore 560080, India}

\author{M. Zuberi}
\affiliation{Department of High Energy Physics, Tata Institute of Fundamental Research, Mumbai-400095, India}

\begin{abstract}
Blazars, a class of highly variable active galactic nuclei, sometimes exhibit Orphan $\gamma$-ray flares. These flares having high flux only in $\gamma$-ray energies do not show significant variations in flux at lower energies. We study the temporal and spectral profile of these Orphan $\gamma$-ray flares in detail from three $\gamma-ray$ bright blazars, 3C~273, PKS~1510-089 and 3C~279 and also their simultaneous broadband emissions.  We find that the variability timescales of the Orphan $\gamma$-ray flares were ($0.96\pm0.28$)~days, ($3.12\pm2.40$)~hr and ($2.16\pm0.72$)~hr, for 3C~273, PKS~1510-089 and 3C~279, respectively. The broadband spectral energy distributions (SEDs) during these flares have been modelled with a leptonic model from two emission regions. This model suggests that Orphan $\gamma$-ray flares might have originated from inverse Compton scattering of relativistic electrons by the seed photons from the broad-line region or dusty torus, which is the first region. While the second broader region, lying further down the jet, could be responsible for X-ray and radio emissions. The possible locations of these emission regions in the jets of the three sources have been estimated from SED modelling. 
\end{abstract}




\section{Introduction}
\label{sec:Introduction}

The broadband spectral energy distributions (SEDs) of blazars show characteristic two broad humps. The low-frequency hump is attributed to synchrotron emission from relativistic electrons, gyrating in the magnetic field of the jet. Blazars can be classified depending on the location of the synchrotron emission peak in the SEDs. The synchrotron emission peak of BL Lacs lies at the optical-ultraviolet-X-ray frequency, while that of FSRQs lies at the infrared frequency. 
The origin of the higher frequency hump in SEDs is possibly inverse Compton (IC) scattering of relativistic electrons by the synchrotron photons (Synchrotron Self Compton, SSC) or the photons external to the jet (External Compton, EC). The external photon field for Comptonization could be from the accretion disk \citep{Dermer1993}, broad-line region (BLR) or dusty torus (DT) \citep{Sikora1994, Blazejowski2000}.  The SSC emission is generally used to explain the high-frequency hump in BL Lacs, while EC emission is preferred for FSRQs due to the presence of seed photons in the circum-nuclear environment of these sources, evident from characteristic emission lines in optical spectra.

Alternatively, it is also possible to produce the higher energy photons in $p-\gamma$ interactions followed by the decay of neutral pions or proton synchrotron process in the hadronic scenario \citep{Mannheim1992, Mannheim1998, Mastichiadis1995, Mastichiadis2005, Mucke2001, Mucke2003, Bottcher2013}. Hadronic interactions have also been studied to investigate the role of AGNs as possible sources of high energy neutrinos \citep{Stecker2013, Murase2014, Tavecchio2015, Petropoulou2015}.

Most of the blazars show simultaneous or nearly simultaneous high flux states at different frequencies, suggesting the common origin of the emissions at these frequencies. However, for some activity periods, time lags between emissions at different frequencies have been observed. This could be due to emissions from different locations along the length of the jet. Some of the high states also show high flux only in one energy band, while having almost constant flux in the other energy bands. In literature, these activities are known as ``Orphan" flares. \cite{MacDonald2015} proposed a two-zone model named ``Ring of Fire" model to explain such flares observed in GeV energies. This model suggests that within the jet a concentrated region of synchrotron photons is produced by a ring-like structure of a shocked non-relativistic sheath of plasma. These ring photons get Compton up-scattered to higher energies by the relativistic electrons of the blob and produce an orphan flare while passing through the ring. The authors explained the orphan flares from six sources, where they decided the orphan $\gamma$-ray flares based on R-band and $\gamma$-ray light curves and reproduced these light curves with their model. Though, the two-zone ``Ring of Fire" model can reproduce the Orphan $\gamma$-ray flares, it can not account for the short timescale variability observed during some of these Orphan $\gamma$-ray flares.  The hadronic synchrotron mirror model was proposed by \cite{bottcher2005, Bottcher2006} for the orphan flare observed in TeV energy from 1ES~1959+650, which constrains the number of relativistic protons in the jet. In the present work, we study the broadband emission during Orphan $\gamma$-ray flares from three sources, 3C~273, PKS~1510-089 and 3C~279. We identify these Orphan $\gamma$-ray flares by examining the multi-waveband light curves from IR to $\gamma$-ray energies.  The periods showing significantly less or no variability in all available and observable energy bands, except $\gamma$-rays, are chosen as the Orphan $\gamma$-ray flares.

Since its discovery \citep{Schmidt1963}, the quasar 3C 273 has been extensively studied covering  radio to $\gamma$-ray frequency. This source is widely used to test and validate the theoretical models of SEDs. Interestingly, this object shows the characteristic feature of the jet of a blazar with superluminal motion, high variability and also that of a Seyfert galaxy having strong blue bump \citep{Malkan1983}, variable emission lines. This sources was studied in $\gamma$-ray by \cite{Rani2013} during the period from September 2009 to April 2010. The authors identified five outbursts from the source, out of which the fifth one was identified as the Orphan flare. From the monthly averaged light curve, the first four flares were inferred as a sub-component of a single flaring event, while the fifth one was seemed to be an independent event.

The FSRQ PKS 1510-089, known for its complex multi-waveband behavior, shows persistent TeV emission. It has been reported in several studies that one-zone model cannot hold for this source \citep{Nalewajko2012, Brown2013, Prince2019a}. The data from \textit{Herschel Space Observatory} PACS and SPIRE instrument, Submillimeter Array and from the instruments used in this work, were used by \cite{Nalewajko2012} in the SEDs of PKS~1510-089. These SEDs have been used in proposing the two-zone model for the jet of this source. The persistent TeV emission from this source from 2012 to 2017 suggested the location of $\gamma$-ray emission region beyond BLR region \citep{Acciari2018}. A time-dependent two-zone model was used by \cite{Prince2019a} to explain the four outbursts of 2015. The authors suggested that the optical-UV and $\gamma$-ray emission originated from a zone within BLR and X-ray emission from the second zone located in DT region. 

At TeV energies, FSRQ 3C~279 was observed by \cite{Aleksic2014}. The authors fitted the June 2011 data with two-zone leptonic model, having one emission region for X-ray to TeV emission, at the inner part of the jet and the other one for optical emission, whose location was derived based on optical polarization data. Moreover, they also suggested the presence of three different emission regions (the third one for radio emission) based on the observed light curves in different energy bands. \cite{Paliya2015} used one-zone leptonic model to explain the IR to $\gamma$-ray emission during March-April 2014 outburst. The authors found the increase in the bulk Lorentz factor as the major cause of this outburst and both, BLR and DT photons, were needed to explain the observed $\gamma$-ray spectrum.

The Orphan $\gamma$-ray flares challenge the one zone emission model which is generally used to explain the simultaneous, near-simultaneous as well as time-averaged (over several months and years) SEDs constructed using multi-frequency observations of blazars. We carefully examine the temporal profile of these Orphan $\gamma$-ray flares and the broadband emissions during these flares from 3C~273, PKS~1510-089 and 3C~279. The multi-wave band data used in this work is mentioned in Section~\ref{sec:Data}. The results of  our analysis  are discussed in Section~\ref{sec:Results}. The Section~\ref{sec:SED} describes the two-zone model used to model the broadband SEDs, which is then followed by further discussion and conclusions in Section~\ref{sec:DC}.

\section{Multi-waveband Data}
\label{sec:Data}

The data were available from space-based telescopes, \textit{Fermi}-LAT, \textit{Swift}-XRT, \textit{Swift}-UVOT and ground-based telescope SMARTS. The analysis procedures adopted for these data are discussed in this section. The data used are from MJD 55250-55290, MJD 54930-54970 and MJD 56730-56770, for 3C~273, PKS~1510-089 and 3C~279, respectively. In addition to the data mentioned in this section, we have also used all the archival data\footnote{https://tools.ssdc.asi.it/SED/} from a web-based tool available at the ASI Science Data Center \citep[ASDC, ][]{Stratta2011}. These data were mainly used to constrain the synchrotron emissions from the sources while modelling the broadband SEDs of three sources.

\subsection{The high energy $\gamma$-ray data}
The high energy data in the energy range from 0.1-300 GeV were obtained from Large Area Telescope onboard \textit{Fermi} satellite\footnote{https://fermi.gsfc.nasa.gov/ssc/data/access/lat/} \citep[$\textit{Fermi}$-LAT,][]{Atwood2009}. It is a pair conversion $\gamma$-ray telescope operational since 2008 August. It scans the entire sky in $\sim$3 hours period due to its large field of view (2.3 $sr$). The \texttt{Fermitools-1.2.23}\footnote{https://github.com/fermi-lat/Fermitools-conda/} were used to analyze the \textit{Fermi}-LAT data. For analysis, the events were extracted from the region of interest (ROI) of 15$^\circ$ centred around source position. To avoid the contamination of background $\gamma$-rays from Earth's limb, the zenith angle cut of 90$^\circ$ was applied. A filter of `(DATA\_QUAL$>$0)\&\&(LAT\_CONFIG==1)' was applied to select good time intervals. The likelihood analysis was performed using $\textit{gtlike}$ \citep{Cash1979,Mattox1996}. The galactic diffuse emission and the isotropic background models, \texttt{gll\_iem\_v07.fits} and \texttt{iso\_P8R3\_SOURCE\_V2\_v1.txt}, respectively with post-launch instrument response function (P8R3\_SOURCE\_V2\_v1) were used in the analysis.

For all three sources, the models used in the analysis are first optimized by performing likelihood fit for the entire period studied in this work over the energy range of 0.1-300 GeV keeping the parameters of all sources within 10$^\circ$ radius free. After the initial fit, we freeze the sources having test statistics (TS; \citet{Mattox1996} less than 25. Also, the parameters of the diffuse galactic model and isotropic background component were left free in the likelihood fit. The sources are modelled using the log-parabola model as given in 4FGL catalogue. This model is used to generate the light curves of different time bins and the SEDs. It is given as, $dN(E)/dE = N_{0} (E/E_p)^{-(\alpha + \beta \log (E/E_p))}$. $E_p$, the value of pivot energy, is fixed to 4FGL catalogue value, and $N_0$ is the normalization parameter. While $\alpha$ and $\beta$ are the spectral index and the curvature parameter of the log-parabola model, respectively.

\subsection{X-ray data}

Soft X-ray data\footnote{https://heasarc.gsfc.nasa.gov/docs/archive.html} in the energy range of 0.3-10.0~keV were obtained from X-ray telescope on board the Neil Gehrels \textit{Swift} observatory \citep[$\textit{Swift}$-XRT,][]{Burrows2005}. These data were analyzed using XRT data analysis software (XRTDAS) distributed within the HEASOFT\footnote{https://heasarc.gsfc.nasa.gov/docs/software/heasoft/} package (v6.26.1). The cleaned event files were generated using \texttt{xrtpipeline-0.13.5}, and \texttt{xrtproduct-0.4.2} was used for the spectral analysis. The source spectrum was extracted from the circle of radius 20 pixels around the source position. While for extracting background spectrum the circular region of 40 pixels in source free region was used. The spectra from different observations were added using \texttt{addspec-1.3.0} and rebinned with a minimum of 20 photons per bin using \texttt{grppha-3.1.0}. The final spectra of all sources were then fitted with an absorbed power-law model ($F(E) = K E^{-\Gamma_x}$) with neutral hydrogen column density ($nH$) fixed at galactic value from \cite{HI4PI2016}. In power-law model, $K$ is the normalization factor in photons keV$^{-1}$ cm$^{-2}$ s$^{-1}$ at 1 keV and $\Gamma_x, $ is the power-law index. The following values of $nH$, 1.69 $\times$ 10$^{20}$ cm$^{-2}$, 7.13 $\times$ 10$^{20}$ cm$^{-2}$ and  2.24 $\times$ 10$^{20}$ cm$^{-2}$, were used for 3C~273, PKS~1510-089 and 3C~279, respectively. Five observations (00035019155, 00035019157, 00035019159, 00035019161, 00035019162) of 3C~279, taken in photon counting mode, were having source counts $>0.5$ counts s$^{-1}$. Hence these observations needed the pile-up correction. To correct for the pile-up effect, the king function, $PSF(r)=[1 + (r/r_c)^2]^{-\beta}$ \citep{Moretti2005}, has been used as given in the standard procedure\footnote{https://www.swift.ac.uk/analysis/xrt/pileup.php}. The level of the pile-up for these observations was 11$\arcsec$-15$\arcsec$. Hence, the annular region having inner and outer radii of 7 and 30 pixels, respectively, was used to extract the source spectrum of these five observations.

\subsection{UV, optical and NIR data}
The optical-UV data\footnote{https://heasarc.gsfc.nasa.gov/docs/archive.html} from \textit{Swift}-UVOT \citep{Roming2005} were used for temporal and spectral studies of the sources. It provides data in three filters in optical band (\textit{V, B} and \textit{U}), and three filters in UV bands (\textit{W1}, \textit{M2} and \textit{W2}). The \texttt{uvotimsum-1.6} was used first to sum the images, and then \texttt{uvotsource-3.3} was used to extract the observed magnitudes. The circular region of 5$\arcsec$ was used for the source region. While for background, the circular region 40$\arcsec$ was used. The observed magnitudes are corrected for galactic extinction of $E_{B-V}$ \citep{Schlafly2011}, which have the magnitudes of 0.0179, 0.101, and 0.029 for 3C~273, PKS~1510-089 and 3C~279, respectively. The corrected observed magnitudes in all six wavebands were then converted into flux using zero-point magnitudes \citep{Poole2008}.

The public archive of Small and Moderate Aperture Research Telescope System \citep[SMARTS,][]{ Bonning2012, Buxton2012} data\footnote{http://www.astro.yale.edu/smarts/glast/3C273hd.php} in the optical and NIR bands were used in the multi-wavelength light curves as well as in the SEDs of 3C~273, PKS~1510-089 and 3C~279. The SMARTS telescope is a part of the Cerro Tololo Inter-American Observatory (CTIO). All the \textit{Fermi}-LAT monitored sources, accessible from Chile, have also been observed by SMARTS in the optical B, V, R bands and NIR J, K bands. The observed magnitudes from SMARTS for all three sources were corrected for the galactic extinction. These corrected magnitudes were then converted into flux density using effective zero points calculated for the Cousins-Glass-Johnson photometric system, and parameters were taken from \cite{Bessell1998}.

\begin{figure}
\centering
\includegraphics[scale=0.5]{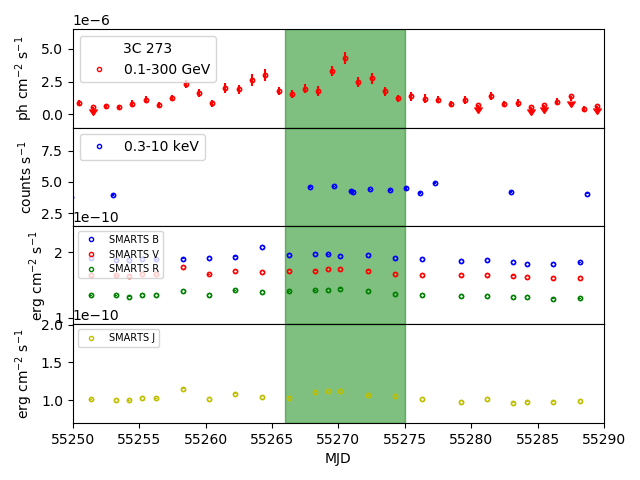}
\caption{The MWLC of 3C 273. Panel-1: One-day bin \textit{Fermi}-LAT light curve; Panel-2: \textit{Swift}-XRT light curve; Panel-3: SMARTS light curves of three optical bands; Panel-4: SMARTS light curve in infrared band.}
\label{fig:mwlc3c273}
\end{figure}

\section{Results}
\label{sec:Results}
The multi-waveband light curves (MWLC) of 3C~273, PKS~1510-089 and 3C~279 help in the clear identification of Orphan $\gamma$-ray flares. 
We examine the temporal evolutions of these Orphan $\gamma$-ray flares by using shorter time bin light curves of these sources and modelling their flare profiles. The details of the MWLC and results of $\gamma$-ray flare modelling are discussed in this section.

\subsection{Multi-waveband light curves}
\label{subsec:MWLC}

\paragraph{3C 273} The source exhibited the Orphan $\gamma$-ray flare with a daily average peak flux of (4.32$\pm$0.45)$\times$10$^{-6}$ ph cm$^{-2}$ s$^{-1}$ on MJD 55275$\pm$0.5. The MWLC from MJD 55250-55290 is shown in Figure~\ref{fig:mwlc3c273}. It can be seen that the source also showed two smaller amplitude flares during MJD 55255 to 55265. At the time of the second flare, the B-band flux seems to be increased marginally. During MJD 55266-55275, the $\gamma$-ray flux increased by a factor of $\sim$2.9, while the fluxes in X-ray, optical and IR energies were almost constant. Hence, this period was chosen as the Orphan $\gamma$-ray flare.

\begin{figure}
\centering
\includegraphics[scale=0.5]{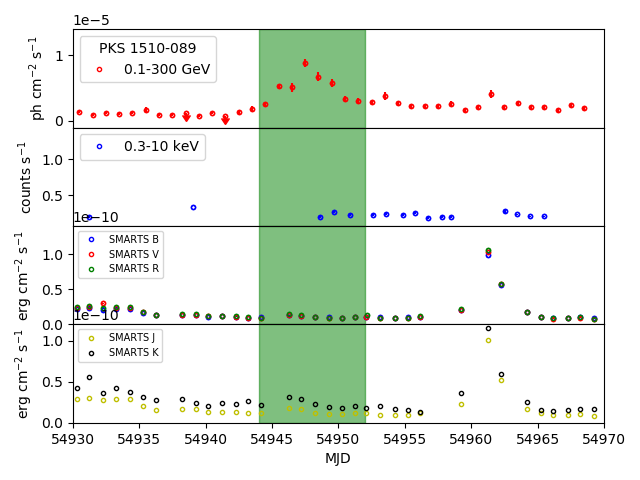}
\caption{The MWLC of PKS 1510-089. The panels same as Figure~\ref{fig:mwlc3c273}.}
\label{fig:mwlcpks1510}
\end{figure}

\paragraph{PKS 1510-089} The Orphan $\gamma$-ray flare from PKS 1510-089 was observed on MJD 54947.5$\pm$0.5 with a daily average peak flux of (8.82$\pm$0.62)$\times$10$^{-6}$ ph cm$^{-2}$ s$^{-1}$. The Figure~\ref{fig:mwlcpks1510} shows the MWLC of this source for the period of MJD 54930-54970. The Orphan $\gamma$-ray period was chosen as MJD 54943-54952, which is marked in the same figure. During this period, $\gamma$-ray flux increased by a factor of $\sim$7.1 without showing much variation at lower energies.  

\begin{figure}
\centering
\includegraphics[scale=0.5]{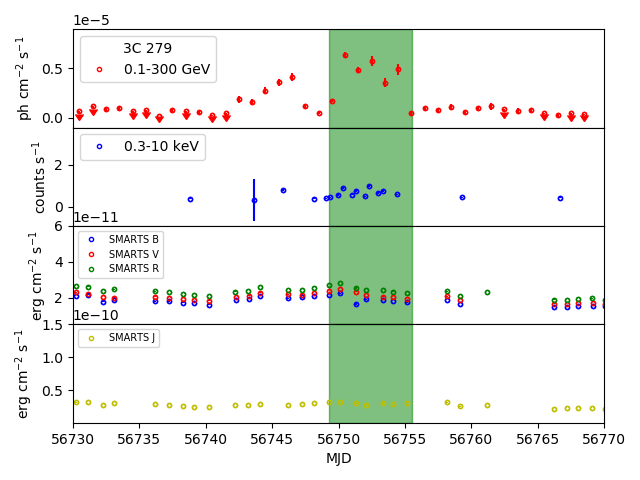}
\caption{The MWLC of 3C 279. The panels  same as Figure~\ref{fig:mwlc3c273}.}
\label{fig:mwlc3c279}
\end{figure}

\paragraph{3C 279} This source showed the Orphan $\gamma$-ray flare on MJD 56750.5$\pm$0.5. It had the daily average peak flux of (6.38$\pm$0.24)$\times$10$^{-6}$ ph cm$^{-2}$ s$^{-1}$. The MWLC from MJD 56730-56770, including the chosen period of Orphan $\gamma$-ray flare is shown in Figure~\ref{fig:mwlc3c279}. This period of Orphan $\gamma$-ray flare lasted for about six days from MJD 56749.25-56755.5, with increased flux by a factor of $\sim$13.2. During this period, the fluxes in other energy bands, shown in Figure~\ref{fig:mwlc3c279} did not show any significant variation.

\subsection{Temporal profile of Orphan $\gamma$-ray flares}
\label{subsec:flarefit}

The three sources studied in this work are among the brightest \textit{Fermi}-LAT blazars, from which fluxes of $>$1$\times$10$^{-5}$ ph cm$^{-2}$s$^{-1}$ above 100 MeV have been observed, during their brightest states. This motivates us to examine the selected flares on shorter timescales. We generate 12 hr binned light curves for 3C~273 and PKS~1510-089, and 6 hr binned light curve for 3C~279. We observed several sub-flaring structures on these timescales during Orphan $\gamma$-ray flares of these sources. We model the light curves showing sub-flaring structures using time-dependent function given by,

\begin{equation}
F(t) = \sum_{i=1}^{n} \{ 2 F_0 \big(\exp^{(t_0-t)/T_r} + \exp^{(t-t_0)/T_d}\big)^{-1} \}_i  + F_b,
\label{Eq:flare}
\end{equation}
where, $t_0$ is the sub-flare peak time, $T_r$ and $T_d$ are the rise and fall time respectively, $F_0$ is the flux at $t_0$, which represents the amplitude of the sub-flare and $F_b$ is the baseline flux. $n$ is the number of sub-flaring components. While modelling these light curves, we successively add the sub-flaring components, until the residue is within $\pm$3 $\sigma$. The residue is defined as the ratio of a difference between model and observed flux over measurement error. A similar method has been adopted while modelling $\gamma$-ray light curves of 3C~454.3 \citep{Abdo2011} and Ton~599 \citep{Patel2018a}.

\begin{table*}
\centering
\caption{The parameters of $\gamma$-ray light curve flare fitting.} 
\resizebox{0.7\textwidth}{!}{
\begin{tabular}{cccccc}
\hline
Source & Name & $F_0$ & $t_0$  & $T_r$ & $T_d$ \\
& & 10$^{-6}$ ph cm$^{2}$ s$^{-1}$ & MJD & day & day \\
\hline
3C 273 	     & s1-1$^{*}$ & 1.0 $\pm$ 0.0   & 55259.30 $\pm$ 0.29 & 1.39 $\pm$ 0.40 & 0.21 $\pm$ 0.19 \\
       	     & s1-2$^{*}$ & 2.0 $\pm$ 0.0   & 55264.40 $\pm$ 0.72 & 1.79 $\pm$ 0.52 & 1.15 $\pm$ 0.62 \\
             & s1-3 & 3.8 $\pm$ 0.0   & 55270.10 $\pm$ 0.27 & 0.96 $\pm$ 0.28 & 1.65 $\pm$ 0.22 \\        
\hline	
PKS 1510-089 & s2-1 & 4.19 $\pm$ 0.55 & 54945.70 $\pm$ 0.21 & 0.39 $\pm$ 0.23 & 1.00 $\pm$ 0.00 \\
	     & s2-2 & 3.69 $\pm$ 0.68 & 54947.20 $\pm$ 0.08 & 0.13 $\pm$ 0.10 & 2.69 $\pm$ 0.70 \\
\hline
3C 279       & s3-1 & 9.50 $\pm$ 0.00 & 56750.40 $\pm$ 0.00 & 0.20 $\pm$ 0.01 & 0.19 $\pm$ 0.01 \\
	     & s3-2 & 7.08 $\pm$ 0.49 & 56752.40 $\pm$ 0.00 & 0.81 $\pm$ 0.07 & 0.37 $\pm$ 0.06 \\
	     & s3-3 & 4.54 $\pm$ 0.88 & 56753.80 $\pm$ 0.00 & 0.20 $\pm$ 0.07 & 0.24 $\pm$ 0.09 \\
	     & s3-4 & 7.32 $\pm$ 0.11 & 56754.60 $\pm$ 0.00 & 0.09 $\pm$ 0.03 & 0.27 $\pm$ 0.04 \\
\hline
\end{tabular}
}
\tablecomments{$^{*}$Not during Orphan $\gamma$-ray flare. The modelled constant background is 6.75$\pm$0.47 10$^{-7}$ ph cm$^{2}$ s$^{-1}$ for 3C 273, 1.84$\pm$0.21 10$^{-6}$ ph cm$^{2}$ s$^{-1}$ for PKS 1510-089 and 2.09$\pm$0.61 10$^{-7}$ ph cm$^{2}$ s$^{-1}$ for 3C 279. }
\label{tab:flarefit}
\end{table*}

\begin{figure}
\centering
\includegraphics[scale=0.5]{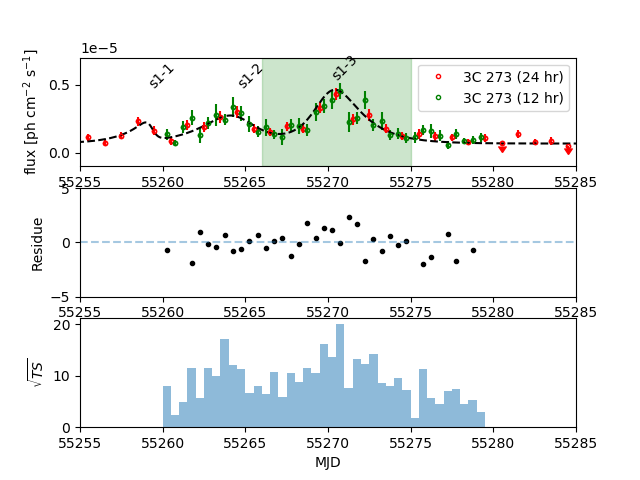}
\caption{Panel-1: 24 hr and 12 hr binned light curve in energy range of 0.1-300 GeV;
Panel-2: Residue of the fit;
Panel-3: Significance of each bin. }
\label{fig:gamma3c273}
\end{figure}

\begin{figure}
\centering
\includegraphics[scale=0.5]{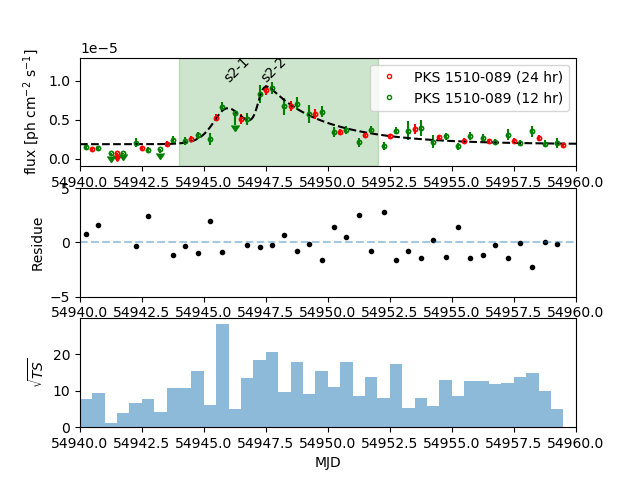}
\caption{The panels are same as Figure~\ref{fig:gamma3c273}.}
\label{fig:gammapks1510}
\end{figure}

\begin{figure}
\centering
\includegraphics[scale=0.5]{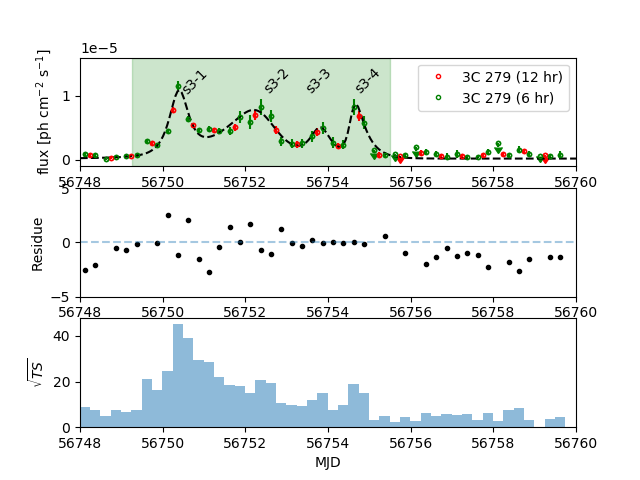}
\caption{Panel-1: 12 hr and 6 hr binned light curves in energy range of 0.1-300 GeV. The bottom two panels are same as Figure~\ref{fig:gamma3c273}.}
\label{fig:gamma3c279}
\end{figure}

The fitted parameters of sub-flaring components for all the three sources are shown in Table~\ref{tab:flarefit}. The modelled $\gamma$-ray light curves for 3C~273, PKS~1510-089 and 3C~279 are shown in Figure~\ref{fig:gamma3c273}, Figure~\ref{fig:gammapks1510} and Figure~\ref{fig:gamma3c279}, respectively. The second panel of each figure shows the residue value of each bin. While the third panel shows the $\sqrt{TS}$ values, which approximately equal the significance \citep{Mattox1996} of flux values. The Orphan $\gamma$-ray flare of 3C~273 was fitted with only one flaring component as both 24 hr, and 12 hr light curves showed a similar pattern. While for PKS~1510 and 3C~279, two and four flaring components, respectively, were needed to reproduce the observed temporal profiles. It can be seen in Table~\ref{tab:flarefit} that most of the components exhibit the fast-rising and slower decay profiles. The fastest rising times during Orphan $\gamma$-ray periods of all three sources were used in modelling the broadband SEDs of these sources. They are (0.96$\pm$0.28) day, (0.13$\pm$0.10) day and (0.09$\pm$0.03) day for 3C~273, PKS~1510-089 and 3C~279, respectively.

\section{Spectral Energy Distribution}
\label{sec:SED}

The Orphan $\gamma$-ray flares studied in present work show sub-flaring structures in $\gamma$-ray energies. As discussed in Section~\ref{subsec:flarefit}, they show these temporal features on a timescale of less than a day. Such structures could be due to the underlying emission mechanism. We study the broadband emissions from these three sources during the Orphan $\gamma$-ray flares by modelling the broadband SEDs. The nine days, ten days and five days averaged broadband SEDs were generated for 3C~273, PKS~1510-089 and 3C~279, respectively. The simultaneous data available during these periods from $\textit{Fermi}$-LAT, \textit{Swift}-XRT, \textit{Swift}-UVOT and SMARTS were used in the SEDs. The spectral analysis results from these observations are mentioned in Table~\ref{tab:SpectralResults}. The archival data available from the ASDC are also used to obtain better constraints on broadband emissions from these sources.

\begin{figure*}
\centering
\includegraphics[scale=0.5]{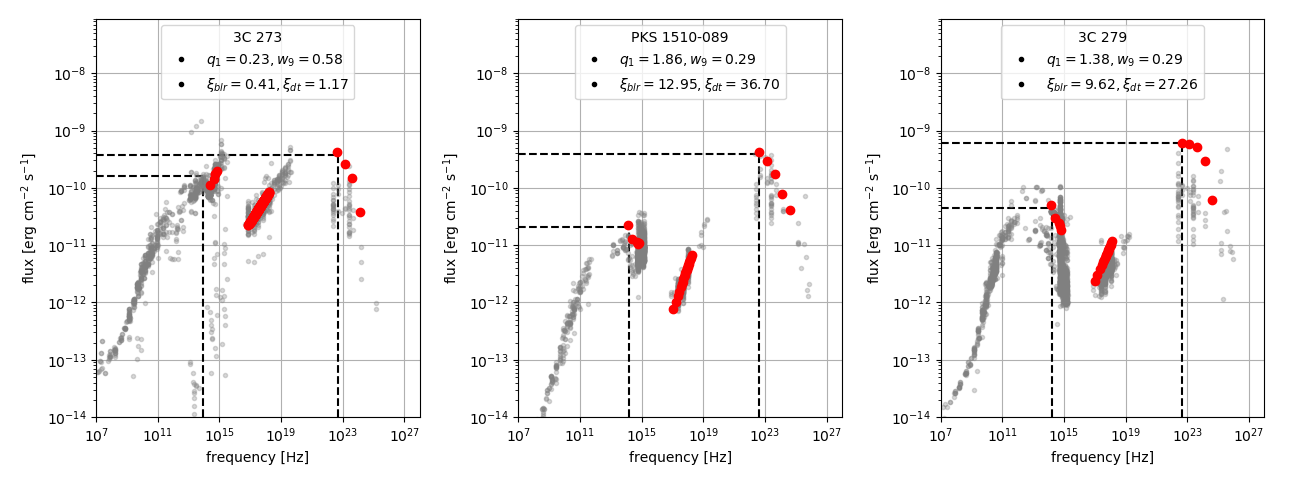}
\caption{The frequencies and fluxes at the synchrotron and IC peaks are shown with dashed lines. }
\label{fig:OneZone}
\end{figure*}

The temporal variability observed in the  multi-waveband light curves, discussed in Section~\ref{subsec:MWLC}, suggests that these emissions might not originate from a single emission region. If these emissions are due to the relativistic electrons of the same emission region, then two physical quantities can be inferred from the broadband SED. The first is the Compton dominance parameter ($q$) and the second is the ratio of synchrotron and IC peak frequencies ($w$). The direct relation between these two parameters, studied by \cite{sikora2009}, depends only on the covering factor ($\xi_{ext}$) and the energy of external photon fields in external frame. This relation, as given by \cite{Nalewajko2012}, can be expressed as follows.

\begin{equation}
\xi_{blr} \approx 0.6 \times \frac{q_1}{w_9^2}, \xi_{dt} \approx 1.7 \times \frac{q_1}{w_9^2 T_3^{5.2}},
\end{equation}

where, $\xi_{blr}$ and $\xi_{dt}$ are the covering factors for BLR and DT, respectively, and $T$ is the temperature of DT, which is assumed as 1000 $K$. The values of these factors are shown in Figure~\ref{fig:OneZone}. It can be seen that, except $\xi_{blr}$ for 3C~273, they are much larger than the values $\sim$0.1-0.3 usually assumed in SED modelling. The large values of covering factors and the variability behaviour observed in the MWLC of these sources, suggest more than one region of broadband emission.

\paragraph{Broadband SED modelling approach} It is mentioned in Section~\ref{sec:Results} that the chosen periods show significant flux variability in $\gamma$-ray energy band on timescale of less than a day. While in the lower energy bands significantly less or no variability was observed during the same periods, compared to $\gamma$-ray energy. This suggests that the $\gamma$-ray emission could be produced inside a compact region while the lower energy emission could be from a comparatively larger region compared to the $\gamma$-ray emission region. Hence, we use the two-zone model having two emission regions (blobs) to reproduce the observed broadband emissions. We assume there are two blobs, blob I having dominant emission in $\gamma$-rays and blob II having dominant emission in optical and X-ray frequencies. We obtain the set of parameters for both the blobs such that the total emission from them constitutes the major part of the observed broadband SEDs.

We have used a publicly available code jetset-1.1.2 code\footnote{https://github.com/andreatramacere/jetset} \citep{Massaro2006, Tramacere2009, Tramacere2011} in this work to model the broadband SEDs. It assumes the spherical blob of size ($R'$), having entangled magnetic field ($B$) inside, is moving with bulk Lorentz factor, $\Gamma$. The blob is viewed at a small angle, $\theta$. This results in a Doppler factor of, $\delta=1/(\Gamma(1-\beta\cos\theta))$. The blob contains the relativistic electrons following power-law distribution. The minimum and maximum electron Lorentz factors are $\gamma_{min}$ and $\gamma_{max}$, respectively. While $p$ is the power-law index. For EC emission, the seed photon fields available are from$-$ (a) the direct emission from the disk \citep{Ghisellini2009}, which has the luminosity of $L_d$ and accreted by the central black hole (BH) having mass, $M_{bh}$, (b) the reprocessed emission in optical-UV frequency from BLR \citep{Donea2003}, and (c) the reprocessed emission in IR frequency from DT, having temperature of T$_{dt}$. The distances of BLR ($R_{blr}$)  and DT ($R_{dt}$) are fixed according to \cite{Ghisellini2009}.

\begin{table*}
\centering
\caption{Results from spectral analysis.}
\label{tab:SpectralResults}
\resizebox{0.8\textwidth}{!}{
\begin{tabular}{lcccccc}
\hline
\textit{Fermi}-LAT: && $\alpha$ & $\beta$ & Pivot energy & $N_0$ & $F_{0.1-300  \text{GeV}}$   \\  
         && & & MeV & ph cm$^{-2}$ s$^{-1}$ MeV$^{-1}$ & 10$^{-6}$ ph cm$^{-2}$ s$^{-1}$   \\  
\hline
        3C 273 && 2.416$\pm$0.071 & 0.062$\pm$0.042 & 279.040 & (3.144$\pm$0.149)$\times$10$^{-9}$ & (2.561 $\pm$ 0.121)\\ 
  PKS 1510-089 && 2.443$\pm$0.022 & 0.063$\pm$0.016 & 743.526 & (2.947$\pm$0.083)$\times$10$^{-10}$& (2.399 $\pm$ 0.047) \\
        3C 279 && 2.223$\pm$0.028 & 0.103$\pm$0.022 & 442.052 & (2.329$\pm$0.074)$\times$10$^{-9}$ & (4.656 $\pm$ 0.120) \\
\hline
\textit{Swift}-XRT: &&&& $\Gamma_x$ &  $K$ & $F_{0.3-10  \text{keV}}$   \\
                    &&&&            &  ph cm$^{-2}$ s$^{-1}$ keV$^{-1}$ & erg cm$^{-2}$ s$^{-1}$ \\
\hline
        3C 273$^{a}$       &&&& 1.555$\pm$0.008 & (2.196$\pm$0.014)$\times$10$^{-2}$ & 1.531$\times$10$^{-10}$ \\
        PKS 1510-089$^{b}$ &&&& 1.213$\pm$0.056 & 8.845$\pm$0.461 & 1.032$\times$10$^{-11}$ \\
        3C 279$^{c}$       &&&& 1.410$\pm$0.003 & (1.356$\pm$0.034)$\times$10$^{-3}$ & 1.252$\times$10$^{-11}$  \\
\hline
\textit{Swift}-UVOT$^{d}$: & w2 & m2 & w1 & uu & bb & vv  \\
&&\\
\hline
     3C 273 & 26.85$\pm$0.572 & - & 23.72$\pm$0.499 & 24.04$\pm$0.546 & - & 13.87$\pm$0.343 \\
     PKS 1510-089 & 1.203$\pm$0.046 & 1.222$\pm$0.044 & - & - & - & -  \\
     3C 279 & 1.363$\pm$0.054 & 1.587$\pm$0.071 & 1.567$\pm$0.067 & 1.945$\pm$0.072 & 2.096$\pm$0.076 & 2.257$\pm$0.095 \\
\hline
SMARTS$^{d}$: && B & V & R & J & K  \\
&&\\
\hline
    3C 273       && 18.06 & 15.88 & 12.71 & 9.427 & -\\
    PKS 1510-089 && 1.073 & 1.043 & 4.680 & 1.287 & 2.256 \\
    3C 279       && 1.892 & 2.187 & 2.492 & 3.064 & - \\
\hline
\end{tabular}
}
\label{tab:fermi}
\tablecomments{$^{a}$Combined spectrum of 00035017063, 00031659002/03/04/05/06/08 observation IDs; $^{b}$Combined spectrum of 00031406001/02, 00031173014 observation IDs; $^{c}$Combined spectrum of 00035019151/52/55/56/57/58/59/60/61/62 observation IDs; $^{d}$\textit{Swift}-UVOT and SMARTS fluxes are in units of 10$^{-11}$ erg cm$^{-2}$ s$^{-1}$. }
\end{table*}

\begin{table}
\centering
\caption{The variability time and corresponding size of $\gamma$-ray emission region}
\begin{tabular}{ccc}
\hline
Source & $t_{var}$ & $R'$\\
       & day & cm \\	       
\hline	       
3C 273       & 0.96 $\pm$ 0.28 & (2.58 - 4.71) $\times$ 10$^{16}$ \\
PKS 1510-089 & 0.13 $\pm$ 0.10 & (0.94 - 7.23) $\times$ 10$^{15}$\\
3C 279       & 0.09 $\pm$ 0.03 & (2.39 - 4.79) $\times$ 10$^{15}$\\    
\hline
\end{tabular}
\label{tab:Size}
\end{table}

\paragraph{The size of emission regions, $\Gamma$ and $\theta$}
The size of blob I is constrained from the  condition that the light crossing time across the blob should be lower than the variability time, $R'\leq c\delta t_{var}/(1+z)$, where the variability time ($t_{var}$) is taken as the fastest flaring time observed during Orphan $\gamma$-ray flares of each source. The values of $\Gamma$ and $\theta$ are used from \cite{Hovatta2009} for all three sources. The  values of $t_{var}$ are mentioned in Table~\ref{tab:Size}. Also, the same values of $\delta$ are used for both the blobs. The values of $\Gamma$, $\theta$ and corresponding $\delta$ are given in Table~\ref{tab:SEDpara}. We make use of the observational fact that the variability at lower frequencies was less than the $\gamma$-ray variability during the periods studied in this work, which suggests that the optical emission comes from a larger region (blob II) than the $\gamma$ rays. We assume the size of blob II about two orders of magnitude larger than blob I.

\begin{table*}[!ht]
\centering
\caption{SED model parameters.}
\resizebox{0.8\textwidth}{!}{
\begin{tabular}{l*{6}{c}}
\hline
Parameter &  \multicolumn{2}{c}{3C 273} & \multicolumn{2}{c}{PKS 1510-089} & \multicolumn{2}{c}{3C 279}\\
          & Blob I & Blob II & Blob I & Blob II & Blob I & Blob II\\
\hline
$R'$ [cm]    & 3.0$\times$10$^{16}$ & 3.08$\times$10$^{18}$ & 6.0$\times$10$^{15}$ & 1.9$\times$10$^{18}$ & 4.5$\times$10$^{15}$ & 1.0$\times$10$^{18}$ \\
$\Gamma$     & 14.0  & 14.0                 & 20.7 & 20.7  & 20.9 & 20.9 \\
$\theta$     & 3.3   & 3.3                  & 3.4  & 3.4   & 2.4  & 2.4  \\
$\delta$     & 16.97 & 16.97                & 16.51& 16.51 & 23.67& 23.67\\
$B$ [G]      & 1.0   & 0.5$\times$10$^{-3}$ & 0.08 & 0.004 & 0.03 & 0.006  \\
$\gamma_{min}$ & 210 & 90                   & 600  & 35    & 500  & 29    \\
$\gamma_{max}$ & 3.0$\times$10$^{3}$ & 2.0$\times$10$^{4}$ & 1.0$\times$10$^{4}$ & 2.2$\times$10$^{4}$ &  1.0$\times$10$^{4}$ & 3.5$\times$10$^{4}$\\
$p$              & 3.4 & 1.82 & 3.4 & 1.8 & 3.3                 & 2.1 \\
$N$ [cm$^{-3}$]  & 2.0 & 2.2 & 700 & 1.15 & 7.5$\times$10$^{3}$ & 6.5 \\
$d$ [cm]             & 6.8$\times$10$^{17}$       & 5.34$\times$10$^{19}$ & 2.0$\times$10$^{18}$  & 3.20$\times$10$^{19}$ & 2.0$\times$10$^{18}$  & 2.38$\times$10$^{19}$ \\
$L_d$ [erg s$^{-1}$] & 4.0$\times$10$^{46}$       & - & 1.0$\times$10$^{46}$       & - & 2.0$\times$10$^{45}$       & -\\
$M_{bh}$ [$M_\odot$] & 5.0$\times$10$^{9}$        & - & $1.0\times$10$^{9}$        & - & 8.0$\times$10$^{8}$        & -\\
$R_{blr}$ [cm]       & (6.0-6.3)$\times$10$^{17}$ & - & (2.0-2.3)$\times$10$^{17}$ & - & (1.4-2.0)$\times$10$^{17}$ &- \\
$\xi_{blr}$          & 0.1 & - & 0.1  & - & 0.1 & - \\
$R_{dt}$ [cm]        & 1.5$\times$10$^{19}$       & - & 6.5$\times$10$^{18}$       & - & 3.5$\times$10$^{18}$       & -  \\
$\xi_{dt}$           & 0.1   & - & 0.3  & - & 0.2  & - \\
$T_{dt}$ [K]         & 1000  & - & 1100 & - & 1000 & - \\
\hline
\end{tabular}
}
\label{tab:SEDpara}
\end{table*}

\begin{table*}
\centering
\caption{The jet power.}
\resizebox{0.7\textwidth}{!}{
\begin{tabular}{*{8}{c}}
\hline
Source & $L_{edd}$ & Blob & $P_e$ & $P_b$ & $P_p$ & $P_t$ & Total power\\
       & erg s$^{-1}$ & & erg s$^{-1}$ & erg s$^{-1}$ & erg s$^{-1}$ & erg s$^{-1}$ & erg s$^{-1}$\\
       \hline
\multirow{2}*{3C 273}       & \multirow{2}*{6.0E+47} & I & 9.53E+42 & 6.61E+44 & 4.99E+42 & 6.75E+44 & \multirow{2}*{2.74E+47} \\
                            &                        & II& 2.15E+47 & 1.74E+42 & 5.81E+46 & 2.73E+47 & \\
&&\\
\multirow{2}*{PKS 1510-089} & \multirow{2}*{1.2E+47} & I & 3.00E+45 & 3.69E+41 & 6.55E+44 & 3.65E+45 & \multirow{2}*{7.96E+46} \\
	                    &                        & II& 5.06E+46 & 9.27E+43 & 2.51E+46 & 7.59E+46 & \\
&&\\
\multirow{2}*{3C 279}       & \multirow{2}*{9.6E+46} & I & 5.27E+45 & 2.98E+40 & 9.39E+44 & 6.21E+45 & \multirow{2}*{8.32E+46} \\
			    &                        & II& 5.22E+46 & 1.04E+44 & 2.47E+46 & 7.70E+46 & \\  	
   \hline
\end{tabular}
}
\label{tab:JetPower}
\end{table*}

\begin{figure}
\centering
\includegraphics[scale=0.6]{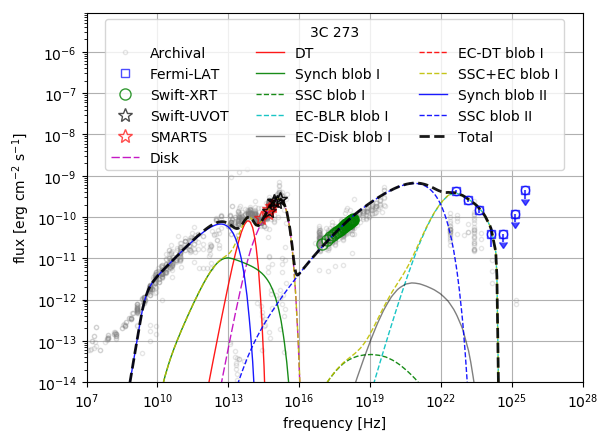}
\caption{Two-zone model fit of 3C 273. The EC-DT component of blob I is below the y-axis scale, hence not visible in the plot.}
\label{fig:sed3c273}
\end{figure}

\paragraph{Broadband SED modelling}
Once we roughly fix the sizes of the blobs, the first step is to constrain the synchrotron and hence the SSC emissions from the sources. These can be constrained using \textit{Swift}-UVOT and SMARTS data if the thermal emissions from the disc are not dominating in this region. For 3C~273 and PKS 1510-089 the thermal emissions from the disc seem to be dominating over non-thermal emissions from the jet. Hence, the archival data from ASDC have been used to constrain the first broad hump of the SED. While for 3C~279, simultaneous \textit{Swift}-UVOT and SMARTS data during Orphan $\gamma$-ray flare helped to constrain the synchrotron emission. Thus the magnetic field $B$ of blob-II is constrained in our two-zone SED modelling. We vary $B$ of blob I along with the electron distribution parameters, $\gamma_{min}$, $\gamma_{max}$ and $p$, in such way that the synchrotron emission and SSC emission for all sources do not exceed the observed IR-Optical-UV emission and X-ray emission, respectively. Thus the magnetic field $B$ in blob I is only marginally constrained. After having the desired electron distribution, the location of blob I ($d$) is varied to get sufficient external photon field density which produces the observed $\gamma$-ray emission by EC process. This constrains the location of blob I in the SED modelling. The values of the fitted parameters of the two-zone model of broadband SEDs of Orphan $\gamma$-ray flares are given Table~\ref{tab:SEDpara}. The modelled SEDs using these parameters are shown in Figure~\ref{fig:sed3c273}, ~\ref{fig:sedpks1510} and ~\ref{fig:sed3c279}, for 3C~273, PKS~1510-089 and 3C~279, respectively.

\section{Discussion and Conclusions}
\label{sec:DC}
The broadband emissions during Orphan $\gamma$-ray flares from three sources, 3C~273, PKS~1510-089 and 3C~279, are modelled using the two-zone leptonic model. These three sources are among the seven brightest \textit{Fermi}-LAT blazars, which occasionally emitted  fluxes of $>$1.0$\times$10$^{-5}$ ph cm$^{-2}$ s$^{-1}$ above 100 MeV. In our model, we infer that the broadband emissions from these sources during Orphan $\gamma$-ray periods, could be from two different regions in their jets, blob I (smaller in size) closer to the jet in BLR/DT region, while the blob II (larger in size) further down the jet. The $\gamma$-ray emission, showing the short timescale (less than a day) variability, could be due to IC scattering of BLR/DT photons by relativistic electrons of blob I. While less or no variability of radio and X-ray fluxes suggests their origin to be of larger size located further down the jet, denoted as blob II in our work. The emissions in the radio and X-ray frequencies are produced by synchrotron and SSC emission of relativistic electrons in blob II. The synchrotron emission from blob II can explain most of the radio data from 3C 273 and PKS 1510-089 and partially the radio data from 3C 279.

The jet powers in different components are calculated from the SED modelling and are mentioned in Table~\ref{tab:JetPower}. It is given by, $P_t=\pi {R'}^2 \Gamma^2 c (u_e'+u_b'+u_p')$. Where, $u_e'$, $u_b'$ and $u_p'$ are the energy densities of relativistic electrons, magnetic field and cold protons (assuming one proton per ten e$^{\pm}$ pairs), respectively, in co-moving frame. The total jet power is computed by summing the jet powers of both the blobs. It is always found to be less than the Eddington Luminosity ($L_{edd}$). The total jet power is $\sim$45$\%$, $\sim$66$\%$ and $\sim$87$\%$ of $L_{edd}$, for 3C~273, PKS~1510-089 and 3C~279, respectively. With accretion efficiency of disk as 0.08, the accretion powers for these sources are 5.0$\times$10$^{47}$ erg s$^{-1}$, 1.25$\times$10$^{47}$ erg s$^{-1}$ and 2.5$\times$10$^{46}$ erg s$^{-1}$, respectively. For 3C~279, the total jet power is larger than the accretion power. This could be possible if there is an extra source to power the jet. This source could be the rotational energy of maximally spinning back hole \citep{Ghisellini2010}.

\begin{figure}
\centering
\includegraphics[scale=0.6]{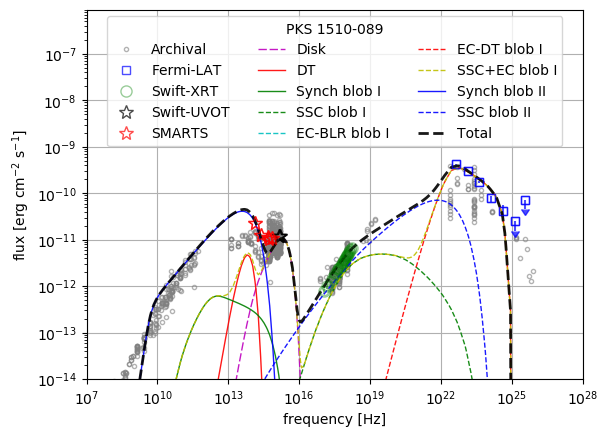}
\caption{Two-zone model fit of PKS 1510-089. In this plot the EC-Disk and EC-BLR component of blob I are below the y-axis scale.}
\label{fig:sedpks1510}
\end{figure}

\begin{figure}
\centering
\includegraphics[scale=0.6]{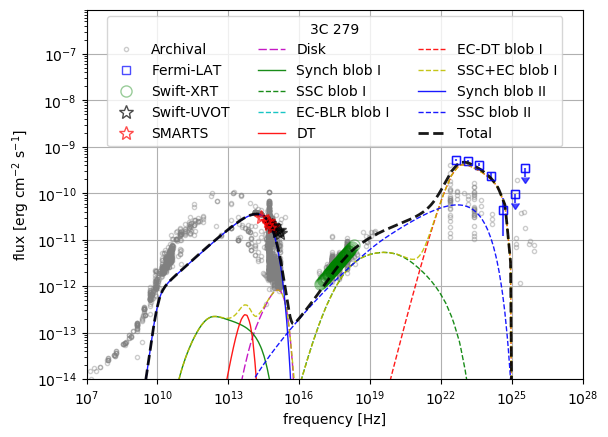}
\caption{Two-zone model fit of 3C 279. The EC-Disk and EC-BLR component of blob I are below the y-axis scale.}
\label{fig:sed3c279}
\end{figure}

The $\gamma$-ray emission is modelled with the EC process with external seed photon field originating from BLR or DT. The intensity of the external seed photon field decreases with distance from the base of the jet. While modelling the SEDs of Orphan $\gamma$-ray flares for the distribution of electrons used in our work, these distances were found to be $\sim$0.22 pc for 3C~273, and $\sim$0.64 pc for PKS~1510-089 and 3C~279. At these distances, the jet cross-sections were found to be $\sim$0.012 pc, $\sim$0.038 pc and $\sim$0.027 pc respectively. They were calculated using the value of distance of blob I from the base of the jet and $\theta$ (assuming $\theta$ as opening angle of the jet) given in Table~\ref{tab:SEDpara}. The emission region of size $R'$ approximately covers the cross-section of the conical jet in case of 3C 273, but it is much smaller than the same in case of PKS~1510-089 and 3C~279. If the jet has the conical geometry beyond parsec scale, then the blob II would be located at distances of $\sim$17.33 pc, $\sim$10.37 pc and $\sim$7.73 pc, for 3C~273, PKS~1510-089 and 3C~279, respectively. In determining the distance of blob II it is assumed that the emission region covers the cross-section of the conical jet.

The optical-UV emission is dominated by the thermal disk component for 3C~273 and PKS~1510-089. While, the disk emission from 3C~279 is much lower than the nonthermal emission from the jet. The X-ray emission for all the three sources were reproduced by SSC process in blob II. While, $\gamma$-ray emission is explained by the EC process happening in blob I. The value of magnetic field is lower in blob II than in blob I, which is consistent with the jet model having conical geometry. After about 15 days, PKS~1510-089 showed another high flux activity but with $\gamma$-ray peak flux comparatively lower than the Orphan flare. However, this later activity showed nearly simultaneous higher flux, about a factor of three than the Orphan flare, in IR-optical bands. This suggests that during the later flare which happened after 15 days, jet contribution at IR-optical energies might be dominating as compared to the Orphan flaring period. The increased $\gamma$-ray emission, which is accompanied by the increase in optical emission indicated their common origin. The Orphan $\gamma$-ray flare of 3C~279 was the brightest among the flares studied in the present work. Also, it lasted for a shorter period compared to the flares from the other two sources. Moreover, PKS~1510-089 and 3C~279 also showed the sub-flaring structures during Orphan $\gamma$-ray period. We also note from Figure~\ref{fig:sedpks1510} and Figure~\ref{fig:sed3c279} that the cumulative X-ray flux from SSC emission of blob-I and blob-II may have slight variation with time due to the contribution from blob-I for the two sources PKS~1510-089 and 3C~279. However, their gamma ray variability is much higher and the flares can be easily distinguished as Orphan gamma ray flares.

The distribution of rise and decay times of the sub-flares is given in Table~\ref{tab:flarefit} in section~\ref{subsec:flarefit}. The rise time is smaller than the decay time for 3C 273. This may happen when the cooling timescale of electrons is longer than their injection/acceleration timescale. In case of PKS 1510-089 the decay times are even longer compared to the rise times. In case of 3C 279 for two sub-flares the rise and decay times are comparable. Nearly equal rise and decay times can be explained by perturbation in the jet or a dense plasma blob passing through a standing shock in the jet region \citep{Blandford1979}. For sub-flare s3-2 the rise time is longer than the decay time and for the sub-flare s3-4 the rise time is shorter than the decay time.  Similar distribution in rise and decay times is also observed for flares of PKS 1510-089 and 3C 454.3 \citep[see,][]{Prince2017, Das2020}.

In this two zone scenario the inner region also produces X-ray emission through SSC process, but its contribution is lower than the outer region. During simultaneous flares in X-ray and $\gamma$-ray frequency, the SSC component from the inner region may dominate over the outer region. 
Thus this two zone model can also be used to explain multi-wavelength flares by adjusting the values of the model parameters.
This model also explains to a large extent the broadband emission at radio frequency from 3C~273 and PKS~1510-089, 
 which is generally not possible to model in a single zone model due to synchrotron self absorption of electrons in the inner part of the jet.  

\acknowledgements
This work has made use of \textit{Fermi} data, obtained from \textit{Fermi} Science Support Center, provided by NASA's Goddard Space Flight Center (GSFC). The data, software and web tools obtained from NASA's High Energy Astrophysics Science Archive Research Center (HEASARC), a service of GSFC is used. The XRT Data Analysis Software (XRTDAS) developed under the responsibility of the ASI Science Data Center (ASDC), Italy has been used in this work. The data from the OVRO 40-m monitoring program \citep{Richards2011} which is supported in part by NASA grants NNX08AW31G, NNX11A043G, and NNX14AQ89G and NSF grant AST-0808050 and AST-1109911 were used. This paper has made use of up-to-date SMARTS optical/near-infrared light curves that are available at www.astro.yale.edu/smarts/glast/home.php. We used \texttt{Fermipy}\footnote{http://fermipy.readthedocs.io} Python package to simplify \textit{Fermi}-LAT analysis.  NG thanks Aditi Agarwal for helpful discussions. DB acknowledges the support of Ramanujan fellowship. SRP thanks A. Tramacere for technical discussion on jetset code.


\bibliography{BlazarPaper}
\bibliographystyle{aasjournal}

\end{document}